\documentstyle[editedvolume]{crckapb} 
\input{psfig.tex}

\begin{opening}
\title{DISTANCES AND AGES OF GLOBULAR CLUSTERS USING HIPPARCOS
PARALLAXES OF LOCAL SUBDWARFS}

\author{RAFFAELE G. GRATTON}
\author{EUGENIO CARRETTA}
\institute{Osservatorio Astronomico di Padova\\
Vicolo dell'Osservatorio 5, I-35012 Padova, ITALY\\
e-mail: gratton@pd.astro.it carretta@pd.astro.it}
\author{GISELLA CLEMENTINI}
\institute{Osservatorio Astronomico di Bologna\\
Via Zamboni 33, I-40126 Bologna, ITALY\\
e-mail: gisella@astbo3.bo.astro.it}
\end{opening}

\runningtitle{DISTANCES TO GLOBULAR CLUSTERS USING SUBDWARFS}

\begin{document}

\begin{abstract}

In this Chapter, first we briefly discuss the impact of Population II and
Globular Cluster (GCs) stars on the derivation of the age of the Universe, and
on the study of the formation and early evolution of galaxies, our own in
particular. The long-standing problem of the actual distance scale to
Population II stars and GCs is then addressed, and a variety of different
methods commonly used to derive distances to Population II stars are briefly
reviewed. Emphasis is given to the discussion of distances and ages for GCs
derived using Hipparcos parallaxes of local subdwarfs. Results obtained by
different authors are slightly different, depending on different assumptions
about metallicity scale, reddenings, and corrections for undetected binaries.
These and other uncertainties present in the method are discussed. Finally, we
outline progress expected in the near future.

\end{abstract}

\section{Introduction}

Since the pioneering work of Baade in the 40's, the study of stellar
populations has been widely recognized as one of the main tool in a variety of
astrophysical problems. Despite their rather large distances and rarity, GCs
can be used as excellent tracers of the oldest stellar populations in
galaxies. Strong constraints can be put on different scenarios proposed for
the formation and early evolution of galaxies by comparing GCs spatial,
kinematical and chemical distributions with those of the underlying field
population.

However, perhaps one of the most powerful uses of GCs is as {\it time
indicators}. These stellar aggregates are believed to have formed within $\sim
1$ Gyr from the Big Bang (see Sandage 1993a), thus the oldest GCs provide very
stringent lower limits to the age of the Universe. It is possible to use GCs
as a clock based on stellar evolutionary principles, to compare with the
cosmological clock given by the Hubble law (the expansion age), once estimates
of the density parameter $\Omega$, and of the cosmological constant $\Lambda$,
are given. The uncertain knowledge of these parameters indicates the
strong advantage of deriving age estimates from nearby objects, exploiting only
properties related to well known stellar evolution theory.

Stars in a GC have all been formed at the same time and, with mainly a few
exceptions, with the same chemical composition; therefore, differences seen
among them are only due to differences in mass. Indeed, the colour-magnitude
diagram (CMD) of a single GC is the observational counterpart of a model
stellar isochrone, i.e. the locus populated by stars with the same chemical
composition and age, but different mass. Stellar models show that the absolute
magnitude of the turn-off (TO) point is a feature that has  reasonably large
sensitivity to age variations, depends on basic stellar properties (mass,
initial chemical composition), but is quite independent of uncertain factors
like mass loss, rotation, mixing, etc., which may have a significant impact on
later evolutionary stages. Therefore ages for GCs can be derived by simply
comparing the observable (absolute magnitude) of the turn-off point with the
corresponding theoretical quantity (bolometric luminosity). The real problem,
in this rather straightforward approach, is in determining the parameters that
enter in the so-called age equation given by stellar models, namely: distance
(which has the largest weight in the error budget, Renzini 1991), reddening,
overall metallicity, and also the detailed pattern of some elemental
abundances, as the $\alpha-$elements. According to all recent studies (see
VandenBerg et al. 1996 for a review), the age-determination problem then
shifts to the distance-scale problem, i.e. to the problem of deriving distance
moduli for GCs with a precision of 0.1 mag or better. This figure translates
into an uncertainty of about 15\% in the derived absolute ages.

Furthermore, GCs host a rather large number of Population II pulsating RR Lyrae
variables. These are generally considered one of the most reliable (i.e.
bright, numerous and well known from a physical point of view) standard candle
within our galaxy and for galaxies in the Local Group. Calibrating the RR
Lyrae distance scale provides an independent estimate (beside that of
classical Pop. I Cepheids) of the distance to the LMC, which is the
traditional {\it first step} in the extragalactic distance ladder. This is a
crucial point, since presently there is a 0.2-0.3 mag difference between the
{\it long} distance scale derived from the period luminosity relation for
Cepheids, and the {\it short} distance scale, derived from the absolute
magnitude of the RR Lyrae variables (see next section).

Coming back to the age determination problem, it is well known that $relative$
ages are easier to obtain than $absolute$ ages. In fact, the displacement of
the colour or magnitude of the TO in different clusters, with respect to a
given standard level, allows a relative comparison that by-passes the exact
knowledge of parameters such as the distance modulus and the reddening. The
final accuracy, however,  heavily depends on which {\it coordinate} of the TO
(horizontal: colour, or vertical: magnitude) is chosen, in the observational
plane, for the comparison. The horizontal approach is more uncertain than the
vertical one, due to the still poor knowledge of the stellar radii. In the
vertical approach, the level of luminosity of the TO is tied to the magnitude
level of the horizontal branch, and both directly rest on the nuclear fuel
(hydrogen or helium) burned in the star core, which is known with great
accuracy from theory (see Stetson et al. 1996 for more details).

Absolute ages with a high degree of precision are more difficult to extract
from observations. Indeed, even if evolutionary models were perfectly adequate
and physically sound, one has to bear in mind that errors of only 0.01 mag in
the observed colours (or of $\sim 0.07$ mag in magnitude, due to the slope of
the TO region) turn into about 1 Gyr uncertainty in the derived age, by itself
about a 10\% error. For the same reason, and in order to maintain the total
error budget below this limit, metal abundances must be known within $0.05
\div 0.1$ dex, and reddening must be determined with the highest precision (one
or two hundredth of mag, if possible). It is easy to understand how, until
recent years, such strict requirements have represented a very difficult
observational test, defying all past efforts to obtain a reliable set of
absolute ages for GCs, at the required level of accuracy.

\section{Methods to derive distances for population II and globular cluster
 stars}

The well-known dichotomy existing between {\it short} and {\it long} distance
scale as derived from old, Population II stars is still an unsolved problem,
in spite of the enormous improvement in distance determinations achieved
with the accomplishment of the Hipparcos mission. The adoption of either of
these two scales directly bears upon the derived model for the Universe, and
for the formation and evolution of the Galaxy.

The most straightforward way to derive distances to Population II stars is to
measure their trigonometric parallaxes; however, GCs are too distant for
present day's detector capabilities. Several {\it indirect} techniques have
therefore been devised, in order to determine the Population II distance
scale. Some of them are based on {\it standard candles}, other exploit
alternative approaches. In the following, the methods most commonly used to
derive distances to Population II objects, as well as some newly proposed
ones, are reviewed, and their virtues and shortcomings briefly discussed.

\subsection{Distances via Standard Candles}

A number of natural {\it standard candles} exist which can be used to
determine accurate distances to Population II objects. They populate different
regions of the HR diagram thus sampling different evolutionary stages of a
Population II star, and include: Red Giant Branch stars, Horizontal Branch
stars (constant stars as well as RR Lyrae variables), Main Sequence Subdwarfs
and White Dwarfs.

\subsubsection{The Red Giant Branch Tip} 

The luminosity in the near-infrared $I$\ band of the brightest stars at the tip
of the Red Giant Branch (TRGB) in GCs changes by less than 0.1 mag in the
metallicity range $-2.2<$[Fe/H]$<-0.7$ (Da Costa \& Armandroff 1990, DA) and,
at fixed metallicity, the $I$\ luminosity of the TRGB varies by less than
$\sim$0.1 mag for ages in the range 7-17 Gyr (Lee, Freedman \& Madore, 1993,
LFM). The observed (dereddened) $I$\ luminosities of the TRGB, transformed to
bolometric luminosities are compared to theoretical TRGB bolometric
magnitudes, (see DA, LFM, and Salaris \& Cassisi, 1997, 1998; SC97, SC98, for
details), and the distance moduli are directly estimated. When calibrated
against models with updated input physics (see SC98 and references therein),
the TRGB method gives distances in good agreement with Hipparcos GCs distance
scale obtained via Main Sequence Fitting. See Madore \& Freedman (this
book) for a thorough discussion of the virtues and shortcomings of this
method.

\subsubsection{The Horizontal Branch}

The absolute magnitude of the HB has often been used to derive distances and
ages for GCs. This is achieved through a number of different
methods:

a) {\it Theoretical HBs}

\noindent 
The absolute magnitude of the HB can be estimated by comparison of the observed
globular cluster HB's to synthetic HB models. This method suffers from several
uncertainties. First, the observed HB morphology is affected from the well
known "second parameter effect" (see Stetson et al 1996 and Fusi Pecci \&
Bellazzini, 1998; for extensive and updated reviews on this problem). Second,
the observed HBs have a non-zero extension in $M_V$\ depending on evolution:
when stars evolve off the zero-age HB (ZAHB) they become brighter and change
colours. Theoretical models usually give the absolute luminosity of the ZAHB.
It is therefore crucial to carefully distinguish among the quantities to
compare: lower envelope of the HB, $M_V(ZAHB)$, mean magnitude of the HB,
$M_V(HB)$, mean absolute magnitudes of the RR Lyrae variables, $M_V(RR)$.
Third, the average absolute magnitude of the HB is a function of metallicity,
the strength of this dependence is not univocally defined though, since
different methods (Baade-Wesselink, Sandage effect, main sequence fitting,
etc.) give a value ranging from 0.16 up to 0.39 for the slope of the
$M_{V}(HB) - {\rm [Fe/H]}$ relation (see Carney et al. 1992, and references
therein). From the theoretical side, HB synthetic models from different
authors generally lead to similar shallow slopes ($\sim$ 0.15-0.20) of the
$M_{V}(HB) - {\rm [Fe/H]}$ relation.

The most recent stellar models that incorporate updated input physics (revised
equation of state, new opacities, improvements of the nuclear cross-sections
and neutrino emission rates), as well as a revised treatment of the stellar
convection and of the helium and/or heavy elements sedimentation (Straniero \&
Chieffi 1997, Caloi, D'Antona \& Mazzitelli 1997, and Cassisi et al. 1998),
tend to give brighter horizontal branch luminosities. When used as standard
candles these new HBs lead to larger distances in agreement with Hipparcos
Main sequence fitting distances. However, other models (VandenBerg 1997) still
tend to provide fainter HBs.

b) {\it Parallax of HB stars}

\noindent 
Gratton (1998) used Hipparcos parallaxes for a sample of field metal-poor HB
stars in order to directly calibrate these standard candles. His sample
consisted of 20 stars with $V<9$, and 2 more stars slightly fainter than this
limit. Three of them are RR Lyrae variables. The mean weighted absolute
magnitude found by Gratton (1998) with this procedure is $M_V=+0.69\pm 0.10$
(at average metallicity [Fe/H]=$-1.41$), and brightens to $M_V=+0.60\pm 0.12$
(at average [Fe/H]=$-1.51$) when HD17072, a suspected first ascent giant
branch star, is eliminated from the sample. This new result is in good
agreement with the methods that favour the {\it short} distance scale,
(statistical parallaxes and Baade-Wesselink), but in conflict with distance
determinations based on globular cluster HB's and Main Sequence Fitting.

c){\it Statistical Parallaxes of RR Lyrae Variables}

\noindent 
Statistical parallaxes of galactic field RR Lyraes give a faint zero point of
the RR Lyrae luminosity calibration, $M_V=0.71\pm 0.12$ at [Fe/H]=$-$1.61
(Layden et al., 1996). This result is confirmed by Popowski \& Gould (1997)
re-analysis of Layden et al. sample ($M_V=0.75\pm 0.13$ at $<$[Fe/H]$> =
-1.61$). Moreover, Fernley et al. (1998a) have used Hipparcos proper motions
for 144 field RR Lyraes and the Statistical Parallax method, as well as the
Hipparcos trigonometric parallax of RR Lyrae itself to estimate the absolute
magnitude of the RR Lyrae stars. They basically confirm Layden et al. results,
and derive $M_V=0.77 \pm 0.15$\ at [Fe/H]=$-$1.53. Similar results are found
by the Hipparcos based statistical parallax analyses of RR Lyrae stars by
Tsujimoto, Miyamoto \& Yoshii (1998) and Gould \& Popowski (1998). The
corresponding distance modulus for the LMC is $(m-M)_{\rm 0,LMC}= <V>_{\rm
0,LMC} - [0.77+0.18({\rm [Fe/H]}+1.53)] = 18.26\pm 0.15$ with  $<V>_0$ =
18.98, and $<$[Fe/H]$< = -1.8$ for the RR Lyraes in the LMC, and adopting a
value of $0.18\pm 0.03$ (Fernley et al. 1998a), for the slope of the
$M_V(HB)$,[Fe/H] relation. All these values still favour the {\it short}
distance scale.

d) {\it Baade-Wesselink for field RR Lyrae variables}

\noindent
The Baade-Wesselink method allows the derivation of the absolute magnitude of
RR Lyrae stars from their intrinsic properties (luminosity, colour and radial
velocity variations). Several groups have used this technique applying it to
field and to a few cluster variables (Liu \& Janes 1990a,b, Jones et al 1992,
Cacciari, Clementini \& Fernley 1992, Skillen et al 1993, Storm, Carney \&
Latham 1994). A general consensus has been reached by these different groups
on a relatively mild slope of the $M_V$,[Fe/H] relation, as opposite to the
steep slope (0.30) found by Sandage (1993b), and on a zero point about 0.3 mag
fainter than derived from classical Cepheids and other methods. In an attempt
to fix the zero-point of the B-W, Fernley (1994) used a different value for
the conversion factor ({\it p}) between observed and true pulsation velocity
getting a zero-point brighter by 0.07 mag, (see also Clementini et al 1995).
However, this only accounts for about 1/3 of the observed discrepancy. More
recently, Feast (1998) and McNamara (1997) have explored different solutions
of the slope and zero-point discrepancy of the B-W results. Feast (1998), from
a compilation of B-W literature data and adopting $M_V$ as independent
variable, derived $M_V = 0.37{\rm [Fe/H]} + 1.13$. When used with Walker (1992)
data, this calibration provides $(m-M)_{\rm 0,LMC}= 18.53$, in agreement with
the classical modulus from Cepheids. However, Feast's paper offers a rather
questionable solution for the slope discrepancy (see also Fernley et al.
1998b), based on the attribution of a zero error to one of the basic
ingredients of the $M_V(HB)$,[Fe/H] relation (i.e. $M_V$), which, on the
contrary, is by far the most uncertain quantity. On the other side, McNamara
(1997) proposes a revision of the B-W results based on the assumption that
optical and near-infrared colours are better temperature indicators than the
$V-K$ index. He uses these colours and the new Kurucz (1993) model atmospheres
to derive the following relation: $M_V = 0.287{\rm [Fe/H]} + 0.964$, that
when applied to the LMC RR Lyraes gives $(m-M)_{\rm 0,LMC} =18.53$ in good
agreement with the Cepheids scale. Clearly much work on the B-W is needed to
definitely assess the correct $M_V(HB)$,[Fe/H] dependence.

\subsubsection{The HB Clump}

Red clump stars are intermediate age (2-10 Gyr) helium core burning stars,
i.e. the metal-rich counterpart of the horizontal branch stars. Red clump
stars were observed in the Galactic Bulge by the OGLE microlensing experiment
(Udalsky et al 1992), and in M31 with the Hubble Space Telescope (Holland,
Fahlman \& Richer 1996, Rich et al 1996). The CMD of the data obtained from
the Hipparcos satellite shows a well determined red clump of
solar-neighbourhood stars at an absolute magnitude of $M_V \sim 0.8$ mag
(Jimenez, Flynn \& Kotoneva 1998). Paczynski \& Stanek, (1998) and Stanek \&
Garnavich (1998), have used the mean absolute magnitude at $I$\ of red clump
stars with parallaxes measured by Hipparcos to derive the distance to the
Galactic Bulge and M31, respectively, and Udalski (1998) has  recently derived
a distance modulus $(m-M)_{\rm 0,LMC}=18.07 \pm 0.12$ by applying the red
clump method to the LMC. Udalski (1998) result, however, is based on a
reddening value for the LMC bar, larger than the commonly adopted value of
$E(B-V)=0.10$\ (Bessel 1991).

The reliability of the red clump method strongly relies on the assumption that
its absolute magnitude is independent of age and chemical composition, and
that the stellar populations in the various systems do not significantly
differ from the solar neighborhood red clump population. These assumptions
have recently been questioned by the model calculations of Girardi et al
(1998), and by Cole (1998) reanalysis of Udalski et al (1998) results for the
LMC. (See Girardi et al, 1998, and Cole, 1998, for a more thorough discussion
of these issues).

\subsubsection{Main Sequence} 

a) {\it Main Sequence Fitting to Isochrones}

\noindent Distances to GCs may be derived by fitting model isochrones to
the observed cluster main sequences. The method is model dependent, and
distances determined by this procedure suffer from the uncertainties still
existing in the equation of state, in the treatment of convection, and in the
transformations from the theoretical $\log{L/L_\odot}-\log{T_{\rm eff}}$\
plane to the observational $M_V-colour$\ plane. Additional concerns are also
represented by the possible existence of phenomena such as the He-diffusion,
core rotation, or problems connected with unconventional physics like
WIMPS (see VandenBerg et al. 1996, for a very extensive discussion of all
these issues).

b) {\it Subdwarf Fitting}

\noindent The simplest technique to derive a distance to a GC is to compare
their main sequence (MS) with a suitable template formed by local, metal-poor
subdwarfs with known trigonometric parallaxes (Sandage 1970). Until the
release of the Hipparcos catalog this procedure mainly suffered due to the the
lack of metal-poor dwarfs in the solar neighborhood with accurate parallaxes
(see VandenBerg et al 1996). With Hipparcos the number of subdwarfs suitable
for main sequence fitting has enormously increased. Moreover, Hipparcos
parallaxes for the local subdwarfs are systematically smaller than the
corresponding ground-based measurements, thus leading, by itself, to longer
distance moduli. In the last year, three different groups employed field
subdwarfs with parallaxes from the Hipparcos mission in order to perform
fitting of globular cluster main sequences: Gratton et al. (1997: G97; 9
clusters), Reid (1997: R97, 1998: R98; 11 clusters) and Pont et al. (1998:
P98; 1 cluster). In total, twelve clusters were analyzed, from the most
metal-poor (M15, M92, and M68), to the most metal-rich ones (M71 and 47 Tuc).
Of the three different studies, P98 essentially confirm the distance, and
hence age, determinations based on subdwarfs parallaxes from ground-based
observations; in contrast, R97, R98, and G97 derive $higher$ distances and
hence $younger$ ages, in overall agreement with each other, and with the
cosmological age for the Universe. This point is further discussed in Section
3 of this Chapter.

\subsubsection{The White Dwarfs Cooling Sequence}

The cooling sequence of white dwarfs provides a faint but theoretically secure
distance ladder, since it is independent of metallicity and age. The
comparison of a globular cluster white dwarf cooling sequence with a template
sequence formed by local white dwarfs with known parallaxes and masses allows
to derive the distance to the cluster. White dwarfs have recently been
observed with HST in 3 GCs: M4 (Richer et al 1995), NGC 6752 (Renzini et al
1996) and 47 Tuc (Zoccali et al 1998, Z98); the two last set of observations
were used to get results favoring a short distance scale. The method is
independent of metallicity and details of the convection theory, but depends
on the observational assumptions related to colours. Furthermore, it assumes
that the calibrating white dwarfs have the same mass of the GC white dwarfs,
an assumption that may be criticized in view of the differences in the age of
the parent populations.

\subsection{The Distance to the Magellanic Clouds}

Walker (this book), gives a detailed review of the methods to derive
distances to the Magellanic Clouds. Here we simply remind that: (i) Feast \&
Catchpole (1997) from a new Period-Luminosity relation for classical Cepheids,
based on the Hipparcos parallaxes of a sample of Galactic Cepheids, derive
$(m-M)_{\rm 0,LMC} =18.70\pm 0.10$. However, Madore \& Freedman (1998), find
that depending on reddening and metallicity effects on the selected sample of
Galactic Cepheids with Hipparcos parallaxes, the LMC distance modulus may
range from 18.44 to 18.57. (ii) van Leeuwen et al (1997) used Hipparcos
parallaxes and infrared photometry of a sample  of Mira variables to establish
the zero point of the Mira period- luminosity relation. They derive a distance
modulus for the LMC of: $(m-M)_{\rm 0,LMC} =18.54\pm 0.18$. (iii)  the "light
echo" of SN1987A leads to distances to LMC ranging from $(m-M)_{\rm 0,LMC}
<18.37\pm 0.04$ (possibly increased to 18.44 for an elliptical shape of the
supernova expansion ring: Gould \& Uza 1998), to $18.58 \pm 0.03$ (Panagia,
Gilmozzi \& Kirshner 1997), and $18.67 \pm 0.08$, Lundqvist \& Sonneborn
(1997).

\subsection{Eclipsing binaries}

Detached eclipsing double-lined spectroscopic binaries have recently been
discovered near the main sequence turn-off of a number of GCs by
Kaluzny et al. (1996,1997). They can be used to directly measure distances to
clusters via a procedure which is similar but simpler than the Baade-Wesselink
for RR Lyrae variables. In practice, the orbital parameters, the 
luminosity-ratios, the size of the orbit, 
and the linear radii of the two components are derived from
the light and radial velocity curves, on a purely geometrical approach. The
latter are combined with the surface brightness of each component inferred
from the observed colours, and distances are so derived. A preliminary 
estimate
of the LMC distance modulus using detached eclipsing binaries gives
$(m-M)_{0,LMC}=18.6\pm 0.2$\ (Guinan et al. 1995). The major shortcoming of
this very promising technique is how to properly derive the surface brightness
from the observed quantities (colours, line ratios etc.).

\subsection{Dynamical models for globular clusters}

Distances to GCs may be derived by comparing proper motion
and radial velocity dispersions within each cluster using King-Michie type
dynamical models. While results for individual clusters derived by this
procedure are affected by large error bars, and depend on cluster dynamical
models, they do not make use of any {\it standard candles} and are totally
independent from stellar evolution models. Rees (1996) gives updated distances
based on this technique for ten GCs. According to Chaboyer et al. (1998)
revision of Rees' results, this method leads to  $M_V(RR)=0.59\pm 0.11$\
(average on 6 clusters). The astrometric distances to GCs might be somewhat
shorter than those derived from subdwarf fittings. Data for a larger number of
clusters would be required to increase the accuracy of the method.

\section{Fitting globular cluster main sequences with local subdwarfs}

In principle, fitting the main sequence of GCs with sequences
of local subdwarfs is a standard {\it one-step} distance calibration which uses 
local
(primary) distance indicators observed in farther objects. As usual in such
distance calibrations, the critical issues are related to the supposed identity
of local calibrators and farther ladders, since it is implicitly assumed that 
(i) the
two samples are extracted from the same parent population (that is: 
the local subdwarf population is supposed to be identical to
the main sequence of GCs); and (ii) the same selection criteria
were applied in selecting objects in the field and in clusters. 
While, naively, the first assumption sounds reasonable in the
present case, it should be critically reviewed to assess its validity. On
the other side, the selection criteria used for local subdwarfs are clearly
very different from those which apply to stars in GCs; hence some
systematic correction must be applied. 

\subsection{Template Main Sequences}

There are two basic difficulties in the fitting procedure. First, the main
sequence is a rather steep relation between colour and magnitude
($4<dV/d(B-V)<7$ in the colour range of interest). Very accurate photometric
data are then required, since any error in the intrisic colours of either
sequences will cause a corresponding (large) error in the derived magnitudes.
Second, due to the sensitivity of the main sequence colour to the chemical
composition the comparison should be made over a restricted range in metal
abundances. Unfortunately, local subdwarfs are rare objects. The most
metal-poor (and likely oldest) GCs have [Fe/H]$<-2$. There are no subdwarfs
with parallax error below 10\% in this metallicity bin, in the Hipparcos
catalogue, and less than 10 with $-2<$[Fe/H]$<-1.5$. Furthermore, the
distribution of stars with metallicity is strongly skewed toward high
abundances, and large errors may be done by misindentifying the mean value
within a bin with the value at mid bin (see discussion in P98).

\begin{figure*}
\vspace{9cm}
\includegraphics{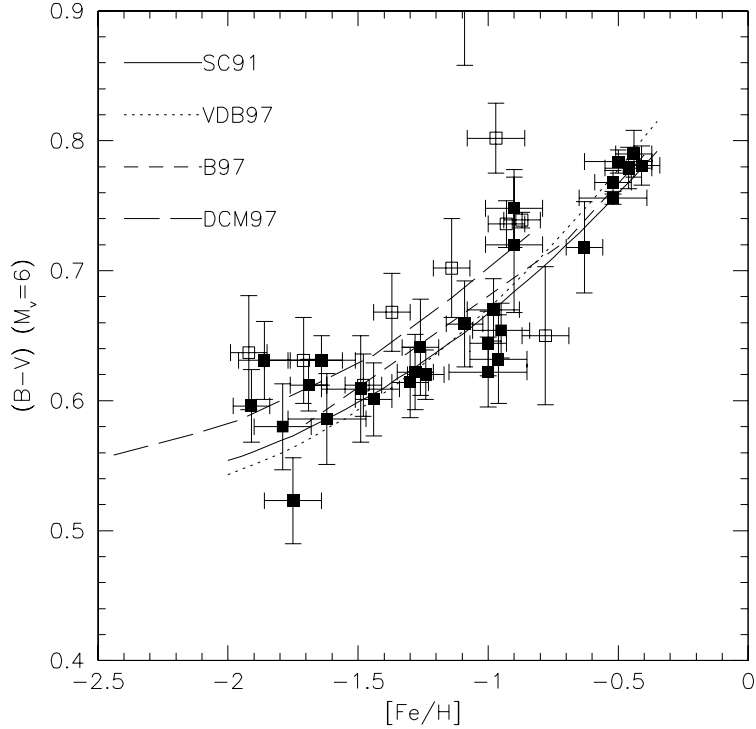}
\medskip
\caption{ Relation between the colour of the main sequence $(B-V)_{Mv=6}$\ and
the metallicity [Fe/H] from local subdwarfs. Filled symbols are {\it bona fide}
single stars; open symbols are known or suspected binaries. Overimposed are
predictions from different isochrone sets (SC91=Straniero \& Chieffi 1991; 
VDB97=VandenBerg 1997; DCM97= D'Antona, Caloi \& Mazzitelli 1997; B97= Bertelli 
et
al. 1997)}
\label{f:fig1}
\end{figure*}

Since unevolved main sequences for systems of different metallicity are
closely parallel to each other in the magnitude range $5.5<M_V<7$\ (G97; this
feature is predicted by models and confirmed by observations of GCs although
the observed slope of the main sequence is slightly shallower than predicted
by most recent models) it is however possible to correct the colour of each
local subdwarf to that the star would have if it had a given absolute
magnitude, and derive a relation between the colour at this absolute magnitude
and metallicity. Once this relation is determined (see Figure~\ref{f:fig1}),
colours of subdwarfs within the magnitude range $5.5<M_V<7$ can be
(empirically) corrected for the metallicity effect, providing template main
sequences for any metallicity within the range covered by the calibrators
(essentially [Fe/H]$>-2$). As shown in Figure~\ref{f:fig1}, various sets of
isochrones (Straniero \& Chieffi 1991; D'Antona, Caloi \& Mazzitelli 1997;
Bertelli et al. 1997; VandenBerg 1997) well reproduce the observed slope of
the colour-metallicity relation (the helium content for all these sets of
isochrones is close to the cosmological value of $Y=0.24$).

\subsection{Physical Biases}

Systematic differences between local subdwarfs and cluster stars might be due
to a different formation mechanism, and/or to dynamical effects of the
environment. For instance, evidences of extra-mixing along the red giant
branch causing anomalous abundances of some elements (C, N, O, Na, Mg, Al) are
found in globular cluster giants, while are much less extreme amongst field
stars (see Kraft 1994 for a review). While the origin of these abundance
anomalies is still unclear, they show that some structural parameter may be
significantly different for stars in different environments. It is generally
assumed that the impact of these differences on the main sequence colours and
magnitudes is small, but they likely have a significant impact in later
evolutionary phases (see e.g. Sweigart \& Catelan 1998).

\subsection{Binarity}

Further concern arises from the contamination due to binary systems. The
presence of a secondary component brightens and reddens the light from the
primary. Hence, main sequences contaminated by binaries would appear brighter
(by as much as 0.75 mag), causing a too long distance scale whether
contamination is in the local subdwarf sample, or a too short distance scale
if the contamination is in the GCs. Binary incidence is likely much higher
amongst field stars than amongst globular cluster stars, since a large
fraction of primordial binaries originally present in GCs are destroyed by
close interactions with other cluster stars during the cluster lifetime. A
large fraction of new binary systems may form in the very dense core of some
GCs; however these regions are not sampled by the accurate and deep
colour-magnitude diagrams considered here. The extensive binary search by
Carney et al. (1994) provided a binary fraction for field metal poor stars
roughly similar to that obtained for Population I stars: likely $\sim 1/2$\ of
stars are binaries. Analogous searches in GCs lead to much lower values ($\leq
20$\%: Pryor et al. 1989; Kaluzny et al. 1998; Rubenstein \& Bailyn 1997;
Ferraro et al. 1997) at least in the outer cluster regions. However, larger
values have been obtained from extensive radial velocity surveys using
Fabry-Perot spectrographs (Fischer et al. 1994; Cote et al. 1994). On the
other side, random star blending (a rather frequent occurrence in the crowded
field of GCs) has the same photometric consequences of physical binarity.

To reduce concern related to binary contamination, the usual approach is to
remove known or suspected binaries from the field stars (see e.g. R97), and to
use modal rather than mean values to identify the main sequence mean loci for
globular clusters (see e.g. Sandquist et al. 1996). However, some residual
undetected binary may still be present among the field stars, and modal
colours may still be somewhat affected by binaries. Unluckily, determination of
accurate binary corrections is difficult, since they depend on both the actual
incidence of binaries, and on the distribution of secondary masses (or
luminosities). A large rough estimate of 0.18 mag was estimated by P98 for the
correction to apply to field stars, on the assumption that half the stars are
binaries, and that the correction for each binary is on average half the
maximum value. This correction was applied to known and suspected binaries, as
well as {\it bona fide} single stars. A much more in depth analysis of this
problem was presented by G97. These authors considered both the average offset
of the colours and the scatter around it, when comparing binaries and {\it
bona fide} single stars. From this comparison, they derived an average
correction for each binary of $0.16\pm 0.05$~mag, and a fraction $p\leq 0.16$\
of undetected binaries in their subdwarf sample (to be added to the 41\% of
stars which are known or suspected binaries). The average binary correction
(in the sense to decrease distances) derived by G97 is of $0.02\pm 0.01$~mag;
this correction must however be applied only to {\it bona fide} single stars
(correction for the whole sample would be $\sim 0.08$~mag). Carretta et al.
(1998, C98) considered the effect of the binary contamination on the mean loci
of GCs. They found that in typical ground-based c-m diagrams, the mean loci
should be $\sim 0.005$~mag redder (with an uncertainty of about 50\%) than the
real single star main sequence, even after the iterative mean procedure has
been adopted. The correction on the cluster distance moduli should then be
$\sim 0.03\pm 0.02$~mag, in the sense to increase distances.

\subsection{Statistical biases}

The statistical correction to parallaxes obviously depend on the sample
selection criteria. We consider here three effects.

Parallaxes are systematically overestimated (and then stars are placed too
close, on average) if stars are selected according to the parallax value
($\pi>\pi_{\rm threshold}$) and/or if weights are given according to the ratio
between the measured parallax and its error ($\pi/\sigma_\pi$). The
corresponding corrections are called Lutz-Kelker corrections from the authors
who first proposed a solution to this problem (Lutz \& Kelker 1973) for a
sample extracted from a population uniformly distributed in space. Lutz-Kelker
corrections are strongly dependent on $\pi/\sigma_\pi$, and are a function of
the parallax distribution of the original population, which is generally not
known. To solve this problem, Hanson (1979) proposed to use the proper motion
($\mu$), since proper motions distribute as $\sim \mu^{1-n}$\ when parallaxes
distribute as $\sim \pi^{-n}$. However, considerable care should be exerted
when Lutz-Kelker corrections are large, because the distribution of parallaxes
may well be different from a simple power law. Best procedure is to use only
stars with good parallaxes ($\sigma_\pi/\pi<0.12$, or similar), because in
this case Lutz-Kelker corrections are small ($<0.2$~mag). On the whole,
Hipparcos provided enough accurate parallaxes for subdwarfs to reduce concern
related to the Lutz-Kelker correction to $<0.02$~mag for the weighted average
of the sample.

If stars have a range of luminosities, a magnitude limited sample would be
biased toward luminous objects (Malmquist bias). The corrections for Malmquist
bias have opposite sign with respect to Lutz-Kelker corrections. Malmquist
bias should be unimportant for unevolved main sequence stars ($M_V>5.5$),
where the intrinsic scatter in luminosity is negligible; while it may
significantly affect the region around the turn-off, where a rather large
intrinsic magnitude range exists, biasing the population toward more evolved
objects. On the whole, the effect should be negligible in the sample
considered by G97 (only unevolved main sequence stars), while it should be
considered (although it is small) for the samples considered by R97 and P98,
who included turn-off stars in their distance determinations. Indeed, P98
derived systematic corrections for Malmquist bias using MonteCarlo simulations.

To enlarge the sample of metal-poor stars with good parallaxes, P98 extracted
stars from the whole Hipparcos catalogue, using selection criteria based on
colours, since most metal-poor stars are expected to be bluer (at a given
absolute magnitude). While these selection criteria added only a few stars
(see C98), the properties of the resulting sample are strongly affected by the
metallicity distribution of the stars in the Hipparcos catalogue, and by the
distribution of errors in colours and magnitudes. Both these quantities are
poorly known. Results obtained from such samples are thus likely to be biased
by stars with colours measured too blue (and hence too faint luminosities).
This problem is overcome by using samples selected according to different
criteria (e.g. proper motions or metal abundances: R97; G97; C98)

\subsection{Errors in stellar parameters}

The largest source of uncertainty in distances to GCs derived via subdwarf
fitting arises from the determination of the stellar parameters. Here, we are
mainly concerned with systematic differences between field subdwarfs and GCs
stars.

\subsubsection{Colours}

Due to the steepness of the main sequence, systematic errors in the photometry
may cause large errors in the derived distance moduli. Ideally, both samples
should be observed on the same photometric system, and, possibly, with the
same instrumental set-up. However, this is made difficult by the large
difference in luminosity (about 10 mag). Indeed, magnitudes and colours for
field stars were obtained with photomultipliers, while GCs are observed with
CCD's. Furthermore, $V-R$\ and $V-I$\ colours in the Johnson-Cousins system
are not available for most field subdwarfs, and colours obtained transforming
data originally taken in other photometric systems (Kron and Johnson $V-R$\
and $V-I$) have a large scatter (Clementini et al. 1998). Since infrared deep
colour magnitude diagrams are not available for most clusters, the only colour
well suited for main sequence fitting is, at present, the $B-V$. Data for GCs
were obtained with CCD's, and transformed to Johnson-Cousins system by
observation of Landolt's standards (Landolt 1983). Albeit considerable care
was devoted to these calibrations, some uncertainty still exists on the derived
transformations, and results for individual clusters may well have rather large
errors (from 0.02 up to 0.04 mag).

\subsubsection{Reddening}

Reddening is a very crucial issue. Again, here we are mainly interested in the
derivation of a uniform scale for both field subdwarfs and GCs. Unfortunately,
a direct comparison of the excitation temperatures of field subdwarfs and
globular cluster main sequence stars is not feasible yet, due to the limits of
the present day's instrumentation (advances are expected with UVES at the
VLT). Nevertheless, some constraints can be derived by comparing the adopted
reddenings with a cosecant law.

Two reddening scales have been used for subdwarfs. R97, G97 and C98 used
reddening estimates from Carney et al. (1994), Schuster \& Nissen (1989), and
Ryan \& Norris (1991). A star-by-star comparison shows that these reddening
estimates are on a uniform scale. The reddenings used by P98 (from Arenou et
al. 1992) are on average 0.016 mag larger (for the 17 stars of P98 Table 2),
leading to a shorter distance scale for globular clusters. On the other side,
all authors used the same scale (from Zinn 1980) for globular cluster
reddenings. (To reduce errors for individual objects, G97 averaged Zinn's
reddening estimates with other most recent values, but these were first put on
the same uniform scale). When compared to cosecant-laws for reddening (like
e.g. that adopted by Bond 1980), we find that GCs and subdwarfs are on a
uniform reddening scale if the height scale of the galactic dust disk is
100~pc when G97, C98 and R97 reddenings are considered; and 40 pc if the
reddenings used by P98 are considered. While the former value is in the middle
of current determinations of the galactic dust scale-height, (50-150 pc:
Lynga 1982, Pandey \& Mahra 1987, Scheffler \& Els\"asser 1987, Spitzer 1978,
Salomon et al. 1979, Burton 1992, Chen 1998), the latter is at the lower
extreme of the admitted range. On the whole, the most appropriate reddening
scale of subdwarfs is still quite uncertain ($\pm 0.015$~mag), with the value
used by G97, C98 and R97 at mid of the admitted range.

\subsubsection{Metal abundances}

A systematic difference of 0.1 dex in the adopted metallicity scale for
subdwarfs and GCs results into an error of $\sim 0.07$~mag (0.03~mag at
[Fe/H]=$-2$, and \ 0.11 mag at [Fe/H]=-1.0) on the distances when main
sequence stars are used, and it is much larger (but of opposite sign) when
subgiants are considered. C98 discussed this point at some length. Clearly, a
uniform abundance analysis should be used (see G97); however, some differential
effect (at $\sim 0.1$~dex) may still be present, since abundances for globular
clusters are derived using giants rather than dwarfs. A substantial progress is
expected in the near future, when UVES at VLT will allow observations of
dwarfs and turn-off stars in GCs.

Metal abundance is not only defined by the Fe abundance, since the
element-to-element ratios may be different from star-to-star. The most
important elements other than Fe are here O and $\alpha$-elements. These element 
are overabundant in most metal-poor stars (Wheeler et
al. 1989; see however King 1997, and Carney et al. 1997 for examples of
metal-poor stars with no excess of O- and $\alpha-$elements) and in globular
clusters (Gratton and Ortolani 1989). While overabundances in the two samples
appear to be quite similar (Carney 1996; Clementini et al. 1998), a slightly
larger excess in cluster stars cannot be excluded. If this were true, the
distance scale would be somewhat underestimated.

\begin{figure}
\centerline{\hbox{\psfig{figure=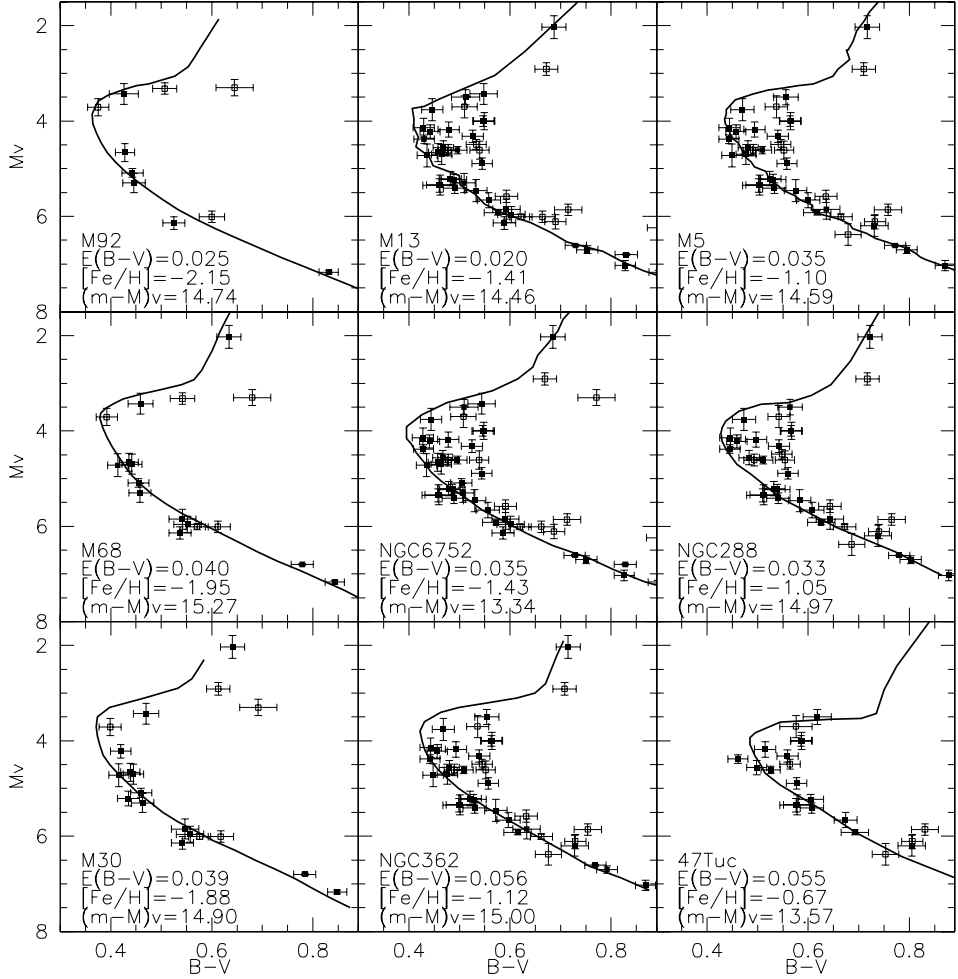,width=13.0cm,clip=}}}
\medskip
\caption{Fits of the fiducial mean loci of the Globular Clusters considered in
C98 with the position of the subdwarfs. Only {\it bona fide} 
single
stars with $5<M_V<8$\ are used in the fits. Values of the parameters
adopted in the analysis (without binary correction) are shown in each
panel}
\label{f:fig2}
\end{figure}

\section{Conclusion}

In this Chapter we have discussed the results provided by the subdwarf-main
sequence fitting method to derive distances and ages for GCs
using Hipparcos trigonometrical parallaxes. Figure~\ref{f:fig2} illustrates
some examples of fitting taken from C98, along with the 
distance
moduli for a number of well observed clusters. Table~1 lists the values
derived for the cluster distance moduli (including corrections for undetected
binaries) by these authors, as well as the ages provided by the 
Straniero \& Chieffi (1991)
isochrones. These are in the middle of the determinations
obtained with various isochrone sets.
\begin{table}
\begin{center}
\caption{GC distance moduli and ages from subdwarf fitting 
(C98)}
\begin{tabular}{lccccc}
\hline \hline
Cluster &$[{\rm Fe/H}]$&E(B-V)&  $(m-M)_V$  &     $M_V(HB)$  &  Age  \\
        &         &       &                 &                & (Gyr) \\
\hline
M92     & $-2.15$ & 0.025 & $14.72\pm 0.07$ & $0.33\pm 0.10$ & 14.8 \\
M68     & $-1.95$ & 0.040 & $15.25\pm 0.06$ & $0.46\pm 0.11$ & 12.3 \\
M30     & $-1.88$ & 0.039 & $14.88\pm 0.05$ & $0.32\pm 0.13$ & 12.3 \\
M13     & $-1.41$ & 0.020 & $14.44\pm 0.04$ & $0.51\pm 0.17$ & 12.6 \\
NGC6752 & $-1.43$ & 0.035 & $13.32\pm 0.04$ & $0.43\pm 0.17$ & 12.9 \\
NGC362  & $-1.12$ & 0.056 & $14.98\pm 0.05$ & $0.45\pm 0.13$ & ~9.9 \\
M5      & $-1.10$ & 0.035 & $14.57\pm 0.05$ & $0.54\pm 0.09$ & 11.2 \\
NGC288  & $-1.05$ & 0.033 & $14.95\pm 0.05$ & $0.45\pm 0.13$ & 11.2 \\
47~Tuc  & $-0.67$ & 0.055 & $13.55\pm 0.09$ & $0.55\pm 0.17$ & 12.5 \\
\hline   
\end{tabular}
\label{t:tab1}
\end{center}
\end{table}

The distance scale obtained by C98 is summarized by the following relation
between the ZAHB absolute magnitude and metallicity:
\begin{equation}
M_V(ZAHB) = (0.18\pm 0.09)({\rm [Fe/H]}+1.5) + (0.53\pm 0.04)
\end{equation} 

Table~\ref{t:tab2} lists the main sources of errors in this distance scale.
The largest contributions come from uncertainties in the reddening scale and
from the adopted metallicity scale. A reasonable guess of the overall
uncertainty (2~$\sigma$) is given by a quadratic sum of the various sources of
errors, this adds to about 0.12~mag, by itself implying a 15\% uncertainty in
derived ages.

Error bars on the ages should also include uncertainties in the stellar
models. Some of the error sources should be interpreted as maximum errors,
while others are standard deviations; furthermore, various error bars are
asymmetric. A Monte Carlo procedure may be devised to give statistically
significative error bars (Chaboyer et al. 1996; G97). Using a similar
approach, C98 best value for the absolute age of the oldest globular cluster
is:
\begin{equation}
{\rm Age} = 12.3^{+2.1}_{-2.5}~{\rm Gyr}
\end{equation}
where the error bars refer to the 95\% confidence range.
\noindent
\begin{table}
\begin{center}
\caption{Systematic errors in GC distance moduli from subdwarf fitting}
\begin{tabular}{lcl}
\hline \hline
Error Source            &   Value        &$\Delta (m-M)$\\
\hline   
random errors           &                & $\pm 0.04$ \\
Lutz-Kelker corrections &                & $\pm 0.02$ \\
binaries                &                & $\pm 0.02$ \\
${\rm [Fe/H]}$-scale    & $\pm 0.1$~dex  & $\pm 0.08$ \\
$E(B-V)$-scale          & $\pm 0.015$~mag & $\pm 0.07$ \\
colour calibration       & $\pm 0.01$~mag & $\pm 0.04$ \\
\hline
total                   &                & $\pm 0.12$ \\
\hline
\end{tabular}
\label{t:tab2}
\end{center}
\end{table}

In order to compare the distance moduli derived from subdwarf fitting with
those from other methods, we list in Table~\ref{t:tab3} the distance modulus
of the LMC obtained by various techniques. Whenever possible, we list
distances originally estimated by the various authors. For some of the methods
based on galactic RR Lyraes where direct estimates were not available we used
$V_0=18.94\pm 0.04$\ and [Fe/H]=$-1.9$\ for the variables in LMC GCs (Walker
1992). For the HB clump method, we give the original distance by Udalski
(1998), as well as the values obtained after correcting for the metallicity
dependence of the clump $M_I$\ magnitude (Girardi et al. 1998), and using a
lower reddening of $E(B-V)=0.10$\ (Bessell 1991). Finally, when considering
the various determinations listed in Table 3, it should be reminded that the
optical depth of the LMC, while small, is not negligible. While some of the
distance estimates refer to the bar (like e.g. those for the Cepheids or for
the HB clump), others are for a few individual objects (SN1987a, GCs with RR
Lyrae) that may well be a few kpc (i.e. up to 10\%) in front or behind the
plane of the LMC.
 
\noindent
\begin{table}
\begin{center}
\caption{True distance modulus to the LMC according to various methods}
\begin{tabular}{lcl}
\hline \hline
Indicator          &$(m-M)_{0,LMC}$& Authors                            \\
\hline
Subdwarf fitting   &$18.54\pm 0.12$& Carretta et al. 1998               \\
                   &               &                                    \\ 
Cepheids           &$18.70\pm 0.10$& Feast \& Catchpole 1997            \\
                   &$18.50\pm 0.07$& Madore \& Freedman 1998            \\
Miras              &$18.54\pm 0.18$& van Leeuwen et al. 1997            \\
SN1987a ring       &$18.37\pm 0.04$& Gould \& Uza 1998 (circular)       \\
                   &$18.44        $& Gould \& Uza 1998 (elliptical)     \\
                   &$18.59\pm 0.03$& Panagia et al. 1997                \\
                   &$18.67\pm 0.08$& Lundqvist \& Sonneborn 1997        \\   
Eclipsing binaries &$18.6 \pm 0.2 $& Guinan et al. 1995                 \\
HB clump           &$18.07\pm 0.12$& Udalski 1998                       \\
                   &$18.31\pm 0.12$& revised by Girardi et al. 1998     \\
                   &$18.43\pm 0.12$& E(B$-$V)=0.10                        \\
HB trig. parallax  &$18.37\pm 0.10$& Gratton 1998                       \\
                   &$18.42\pm 0.12$& eliminating HD17072                \\
Stat. parallax     &$18.29\pm 0.12$& Layden et al. 1996                 \\ 
                   &$18.26\pm 0.15$& Fernley et al. 1998a               \\ 
Baade-Wesselink    &$18.26\pm 0.04$& Clementini et al. 1995             \\
                   &$18.34\pm 0.04$& Clementini et al. 1995 (p=1.38)    \\
                   &$18.53\pm 0.05$& McNamara 1997                      \\
GC Dynamical models&$18.44\pm 0.11$& Rees 1996, revised by Chaboyer et al 1998\\
\hline
\end{tabular}
\label{t:tab3}
\end{center}
\end{table}

We conclude by giving a few indications for future observations aimed at
refining the present distance determinations. Current available instrumentation
(both ground based and the HST) may be used to improve the photometric
calibrations, that in various cases may be uncertain. It is particularly
important to repeat the observations for M92, which yields by far the largest
age, and it is then the best candidate as the oldest GC. Determination of
consistent reddening and metal abundance scales for GCs and local subdwarfs is
a crucial step. Significant progresses in this respect are expected using UVES
at VLT2, which will get high resolution, high $S/N$\ spectra of cluster main
sequence stars. High precision (error $\sim 1$~mas) ground-based parallaxes on
the Hipparcos absolute scale may be obtained for some very metal-poor
([Fe/H]$<-2$) subdwarfs with $6<M_V<8$\ and $V>11$. Finally, trigonometric
parallaxes of GC stars will be directly measurable when GAIA will be launched.

{}

\end{document}